\documentclass{aa}  

\usepackage{graphicx}
\usepackage{txfonts}
\usepackage{gensymb}
\usepackage{natbib}

\newcommand{\eg}{{e.g.}}
\newcommand{\ie}{{i.e.}}
\newcommand{\new}[1]{{ #1}}

\begin{document}

   \title{Multiwavelength identification of millisecond pulsar candidates in the Galactic bulge}

   \subtitle{}

   \author{J. Berteaud
          \inst{1,2,3,4}
          \and
          F. Calore\inst{1}
          \and
          M. Clavel\inst{2}
          \and
          J. Marvil\inst{5}
          \and
          S. Hyman\inst{6}
          \and
          F. K. Schinzel\inst{5}
          \and
          M. Kerr\inst{7}
          }

   \institute{LAPTh, CNRS, USMB, F-74940 Annecy, France\\
              \email{berteaud@lapth.cnrs.fr}
         \and
             Université Grenoble Alpes, CNRS, IPAG, F-38000 Grenoble, France
         \and
                University of Maryland, Department of Astronomy, College Park, MD 20742, USA
         \and
                NASA Goddard Space Flight Center, Code 662, Greenbelt, MD 20771, USA
         \and
                        National Radio Astronomy Observatory, P.O. Box O, Socorro, NM 87801, USA
         \and
                        Dept. of Engineering \& Physics, Sweet Briar College, Sweet Briar, VA 24595, USA
         \and
                        Space Science Division, US Naval Research Laboratory, Washington, DC 20375, USA\\
             }

   \date{Received XXXXX, 2023; accepted YYYYY, 202Z}

 
  \abstract
   {The existence of a population of millisecond pulsars in the Galactic bulge is supported, along with other evidence, by the \textit{Fermi} GeV excess, an anomalous $\gamma$-ray emission detected almost 15 years ago in the direction of the Galactic center. However, radio surveys searching for pulsations have not yet revealed bulge millisecond pulsars.}
   {Identifying promising bulge millisecond pulsar candidates is key to motivating pointed radio pulsation searches. Candidates are often selected among steep-spectrum or polarized radio sources, but multiwavelength information can also be exploited: The aim of this work is to pinpoint strong candidates among the yet unidentified X-ray sources.}
   {We investigated the multiwavelength counterparts of sources detected by the \textit{Chandra} X-ray observatory that have spectral properties expected for millisecond pulsars in the Galactic bulge. We considered that ultraviolet, optical, and strong infrared counterparts indicate that an X-ray source is not a bulge pulsar, while a radio or a faint infrared counterpart makes it a promising candidate.}
   {We identify a large population of more than a thousand X-ray sources without optical, ultraviolet, or strong infrared counterparts. Among them, five are seen for the first time in unpublished radio imaging data from the Very Large Array. We provide the list of promising candidates, for most of which follow-up pulsation searches are ongoing.}
   {}


   \maketitle
%

\section{Introduction}

Among the zoology of objects in our Galaxy, pulsars stand out as peculiar sources. These highly magnetized and fast rotating neutron stars are characterized by a very stable pulsed emission, which enables a sure identification. They are multiwavelength emitters, notably in $\gamma$ rays \citep[see \eg,][]{2013ApJS..208...17A,2023ApJ...958..191S}, X rays \citep[see \eg,][]{2018ApJ...864...23L,2020MNRAS.492.1025C}, and radio \citep[see \eg,][]{2005AJ....129.1993M}.

Pulsars are fantastic probes in many physics domains. The dispersion observed in their pulsed radio emission informs us about the properties of the interstellar medium and is used to create models of the Galactic density of free electrons \citep[see \eg,][]{2017ApJ...835...29Y}. Simultaneous measurements of the mass and radius of pulsars provide constraints on the equation of state of cold dense matter \citep[see \eg,][]{2022NatRP...4..237Y}. Pulsars are also unique laboratories for studying physics and testing theories of gravity in extreme conditions (see, \eg, \cite{2016ApJ...818..121P, 2023ApJ...959...14T} and \cite{1999ApJ...514..388W, 2021PhRvX..11d1050K} for pulsars in binary systems). Recently, pulsar timing arrays have provided evidence for a stochastic gravitational wave background \citep{2023ApJ...951L...8A, 2023A&A...678A..50E, 2023ApJ...951L...6R, 2023RAA....23g5024X}.

The majority of the known pulsar population is local and found within a few kiloparsecs of Earth. Very little is known about the pulsar population toward the Galactic center (GC), making this region, especially the one around the supermassive black hole Sgr A$^*$, a very interesting place to search for new pulsars. Indeed, the GC is a massive-star formation site \citep{2009msfp.book...40F}, harboring neutron star progenitors, some pulsar wind nebulae candidates \citep{2008ApJ...673..251M}, numerous point-like X-ray sources \citep{2009ApJS..181..110M}, and dense cusps of X-ray binaries, possibly including millisecond pulsars (MSPs), that is, pulsars with rotation periods < 30 ms \citep{2018Natur.556...70H,2021ApJ...921..148M}. Moreover, theoretical studies and multiwavelength observations suggest that a large population of pulsars exists in the vicinity of Sgr~A$^{*}$, as well as on larger scales \citep{2004ApJ...615..253P}. The unexplained, spatially extended, diffuse $\gamma$-ray emission detected by the \textit{Fermi}-LAT, the so-called \textit{Fermi} GeV excess, can be interpreted as the cumulative emission from a population of MSPs, too faint to be detected as individual $\gamma$-ray sources and with number density strongly peaked at the GC (for a review, see \cite{2020ARNPS..70..455M} and references therein). An alternative, compelling, hypothesis claims that the \textit{Fermi} GeV excess is caused by dark-matter annihilation in the GC. Therefore, discovering MSPs in the Galactic bulge can also help to constrain the contribution of dark matter to the excess and its properties. Recent independent $\gamma$-ray analyses have confirmed an at least partial stellar origin of the \textit{Fermi} GeV excess, strengthening the case for the existence of an abundant population of MSPs in the Galactic bulge \citep{2021PhRvL.127p1102C, 2020PhRvL.125x1102L,2022PhRvD.105f3017M,2021PhRvD.104l3022L}.

It has been shown that current radio timing surveys are not sensitive enough to unveil the bulge MSP population \citep{2016ApJ...827..143C}. Deep targeted observations are therefore needed. Pulsar candidates are usually selected among steep-spectrum (index $\alpha < -2$) and/or highly polarized radio sources \citep[see, \eg][]{2019ApJ...884...96K, 2019ApJ...876...20H, 2022A&A...661A..87S}. Indeed, pulsars are part of the sources with the steepest radio indices \citep{2000ApJ...529..859K} and the highest circular polarization fraction \citep{2018MNRAS.478.2835L}. Nonetheless, the mean radio index of MSPs is $-1.77\pm 0.74$\footnote{Computed from the 63 MSPs in the Australia Telescope National Facility (ATNF) pulsar catalog \citep{2005AJ....129.1993M} with 400 and 1400 MHz known fluxes.} and high circular polarization is not universal among pulsars \citep{1998MNRAS.300..373H}. Thereby, standard MSP candidate identification methods focus on a peculiar part of the whole population.

\begin{figure*}
    \includegraphics[width = \hsize]{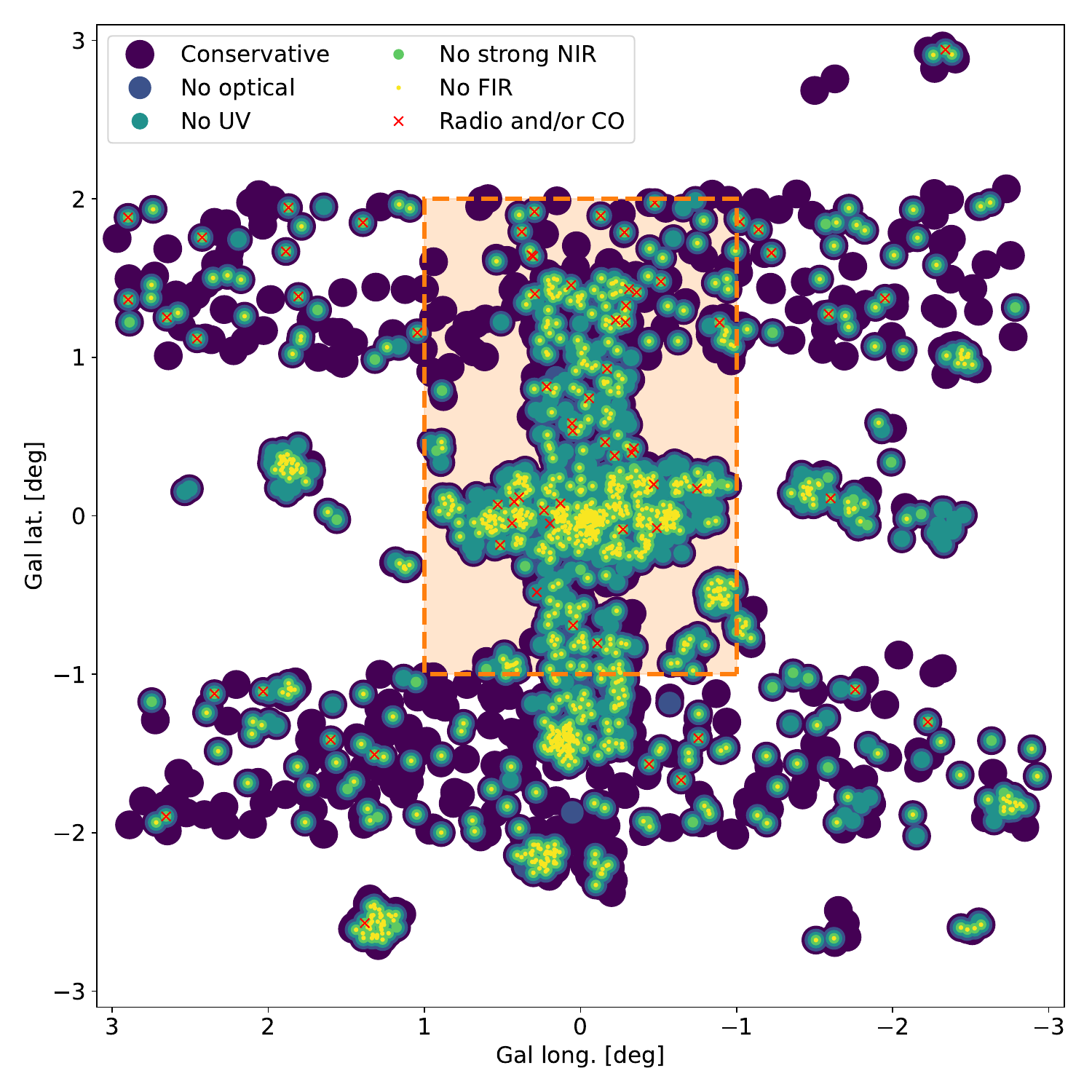}
    \caption{Positions of MSP candidates after successive selections: conservative selection of \cite{2021PhRvD.104d3007B} (dark blue) after removing those with optical (Section \ref{sec:cand_opt}, blue), UV (Section \ref{sec:cand_uv}, teal), strong NIR (Section \ref{sec:cand_nir}, green), and FIR (Section \ref{sec:cand_fir}, yellow) counterparts. Some of these sources have a potential VLA (radio) counterpart and/or are compact-object (CO) candidates (red crosses). \new{Sources from the conservative selection of \cite{2021PhRvD.104d3007B} for which we found an optical counterpart therefore appear as single dark blue circles; those without an optical counterpart but with a UV one appear as a blue circle with a dark blue ring, etc.} The shaded region shows the approximate coverage of the VLA mosaic (orange). The spatial distribution of the candidates is correlated with the sensitivity threshold of \textit{Chandra}.}
    \label{fig:cand_lb}
\end{figure*}

\cite{2021PhRvD.104d3007B} simulated the Galactic population of disk and bulge MSPs, together with their $\gamma$- and X-ray emissions, assuming the bulge component to be the only thing responsible for the \textit{Fermi} GeV excess. This study has assessed the sensitivity of existing X-ray telescopes to this population for the first time and demonstrates that about a hundred bulge MSPs could have been detected in past observations of the \textit{Chandra} X-ray Observatory in our region of interest (ROI; $|l|, |b| < 3\degree$), while the contamination from disk MSPs is negligible. \cite{2021PhRvD.104d3007B} conservatively selected $3158$ bulge MSP candidates\footnote{3153 in our previous work; the difference is due to the use of the \textit{Chandra} Source Catalog (CSC) equatorial -- which are more precise -- instead of the CSC Galactic coordinates when looking for counterparts.} among sources in the second release of the \textit{Chandra} source catalog \citep{2010ApJS..189...37E}, according to their X-ray spectral behavior and their distance, excluding (too) soft sources and those either too close or too far to be in the bulge. Their positions are shown in Figure \ref{fig:cand_lb}. The large number of MSP candidates compared to the expected number of detections does not allow us to rule out the MSP interpretation of the \textit{Fermi} GeV excess and motivates further investigations.

The classification of X-ray sources through multiwavelength counterparts is an active field of research (see, \eg, \cite{2022A&A...657A.138T}), but it often does not include a proper category for pulsars. 
In this paper, we present a novel method for the identification of (millisecond-)pulsar candidates from multiwavelength data (starting from the X rays) free from standard polarization and radio spectral index criteria. Our goal is to highlight a small subset of sources that is promising enough for follow-up studies. To this end, we exploited the typical multiwavelength pulsar emission. 

In Section \ref{sec:cat_obs}, we describe the catalogs and observations that we used to associate our X-ray MSP candidates with their potential counterparts and how we did the associations. Section \ref{sec:mw_msp} presents our results and highlights the candidates that we selected for follow-up observations. Finally, in Section \ref{sec:msp_mw_sumr}, we summarize our results and discuss our method and future prospects.

\section{Cross-matches: Catalogs and observations}
\label{sec:cat_obs}

To investigate the multiwavelength counterparts of the \textit{Chandra} MSP candidates, we used published catalogs in the optical, ultraviolet (UV), infrared (IR), and radio, as well as radio observations obtained with the Very Large Array (VLA), which are presented in this paper for the first time.

\subsection{Cross-matches}
\label{sec:crossmatch}

The association between \textit{Chandra} and multiwavelength sources is not, in most of the cases, known a priori. We performed geometrical cross-matches between our \textit{Chandra} candidates and multiwavelength sources. A positive match satisfies
\begin{equation}
        \theta_\mathrm{sep} < \sqrt{\texttt{err\_ellipse\_r0}^2 + \theta_\mathrm{MW}^2},
    \label{eq:pos_cm}
\end{equation}
where $\theta_\mathrm{sep}$ is the angular separation between a \textit{Chandra} source and a multiwavelength source, \texttt{err\_ellipse\_r0} is the major radius of the 95\% confidence level (CL) position error ellipse of the \textit{Chandra} source and $\theta_\mathrm{MW}$ is the error on the position of the multiwavelength source at the same CL. \new{The CSC provides error measurement at the 95\% CL only. Therefore, for consistency, we use this CL throughout the paper.} We note that this cross-match method inevitably leads to spurious associations. Our primary goal here, however, is not to formally associate \textit{Chandra} sources with their multiwavelength counterparts, but rather to identify those having potential association(s). Depending on the emission intensity and spectral domain of the counterpart, our associations are either excluding or accepting candidates. Our cross-matching procedure is a trade off between purity and completeness, that is, we reject sources to increase the proportion of true MSPs in our sample, but the latter might not be complete because some MSPs could be mistakenly rejected. We note that such trade off is commonly applied in pulsar searches (see, \eg, \cite{2022A&A...661A..87S}, who furthermore used a constant cross-match radius of 10 arcsec around radio sources). A second parameter of the \textit{Chandra} sources that will be useful for some of the cross-matches is the energy flux recorded in the \textit{Chandra} broad band (0.5--7 keV), \texttt{flux\_aper90\_b}.

\subsection{Catalogs}

\subsubsection{Optical}
\label{sec:cat_opt}

The \textit{Gaia} mission \citep{2016A&A...595A...1G} has mapped the entire celestial sphere for nearly a decade and provides precise measurements for more than 1 billion stars down to magnitudes of $\sim$20 in the G band (330--1050 nm). The latest \textit{Gaia} data release contains positions and G-band magnitudes for 1.8 billion sources. We associated our \textit{Chandra} MSP candidates with their potential \textit{Gaia} counterparts following the procedure described in Section \ref{sec:crossmatch}, neglecting the error on the optical sources by setting
\begin{equation}
    \theta_\mathrm{MW} = 0
    \label{eq:sep_nir}
\end{equation}
in Equation \ref{eq:pos_cm}. The results of the X-optical associations are presented in Section \ref{sec:cand_opt} as well as the resulting exclusions.

\subsubsection{Ultraviolet}

For the UV associations, we made use of the latest release of the XMM-OM Serendipitous Ultraviolet Source Survey catalog, XMM-OM-SUSS5.0\footnote{\url{https://www.cosmos.esa.int/web/xmm-newton/om-catalogue}} \citep{2012MNRAS.426..903P}. The catalog contains more than 8 million entries, corresponding to almost 6 million sources detected by the Optical Monitor (OM) on board the X-ray Multi-Mirror Mission spacecraft, XMM-Newton, between 180 and 361 nm. About 280~000 entries fall in our ROI, but the sky coverage of XMM-OM, similar to \textit{Chandra}'s one, is patchy. Hence, some MSP candidate positions were not observed and cannot reveal any UV counterpart. However, the UV survey covers almost entirely the region with $|l| < 1\degree$ and $-2\degree < b < 1.5\degree$, which contains the majority of our candidates. We followed the cross-match procedure described in Section \ref{sec:crossmatch} with
\begin{equation}
    \theta_\mathrm{MV} = 2 \times \texttt{POSERR}
,\end{equation}
where \texttt{POSERR} is the 68\% CL uncertainty on the position of the UV source. We assumed that \texttt{POSERR} has a Gaussian distribution, so that $2 \times \texttt{POSERR}$ is the 95\% CL uncertainty on the position. The results of the X-UV associations are presented in Section \ref{sec:cand_uv} together with the resulting exclusions.

\subsubsection{Near-infrared}

We performed the \textit{Chandra}-near IR (NIR) associations with three different surveys. The first one is the Two Micron All Sky Survey \citep[hereafter 2MASS,][]{2006AJ....131.1163S}, which has detected more than 500 million sources over the entire sky, in the J (1.235 $\mu$m), H (1.662 $\mu$m), and K$_\mathrm{s}$bands (2.159 $\mu$m). The magnitude of the detected sources ranges between --4 and 16 mag, with an astrometric accuracy better than 300 mas.

Second, we used the last release\footnote{\url{https://cdsarc.cds.unistra.fr/viz-bin/cat/II/348}} of the VISTA Variables of the Via Lactea survey \citep[hereafter VVV,][]{2010NewA...15..433M}, which has observed the inner Galaxy and identified about half a billion of sources down to K$_\mathrm{s}$ magnitudes of $\sim$17\,mag. Sources without a \texttt{Ksmag1}, the aperture-corrected K$_\mathrm{s}$ magnitude for the smallest source diameter (1 arcsec), in the VVV catalog, were ignored in this work.

Finally, we used the catalog of the United Kingdom Infrared Telescope Deep Sky Survey of the Galactic plane\footnote{\url{https://cdsarc.cds.unistra.fr/viz-bin/Cat?II/316}} \citep[hereafter UKIDSS,][]{2008MNRAS.391..136L}, which contains about half a billion sources. The median 5$\sigma$ depth of the survey is better than $\sim$18\,mag in all three bands (J, H, and K$_\mathrm{s}$). The UKIDSS catalog provides two measurements of the K$_\mathrm{s}$ magnitude, \texttt{Kmag1} and \texttt{Kmag2}, corresponding to two different epochs. Sources with no K$_\mathrm{s}$ measurement were ignored.

All three catalogs presented above cover our whole ROI. As for the \textit{Gaia} objects, we have neglected the positional error of the NIR sources in the cross-match procedure (equation \ref{eq:sep_nir}). The exclusion or acceptance of candidates resulting from these associations depends on the X-ray and the NIR emissions; the results are presented in Section \ref{sec:cand_nir}.

We also looked for counterparts in the 2MASS Extended Catalog \citep[2MASX,][]{2000AJ....119.2498J}. We identified the MSP candidates overlapping 2MASX sources by replacing Equation \ref{eq:pos_cm} by
\begin{equation}
        \theta_\mathrm{sep} < r_\mathrm{ext},
    \label{eq:pos_cm_extended}
\end{equation}
in the cross-match procedure, with $r_\mathrm{ext}$ being the radius of the 2MASX source. Results are presented in Section \ref{sec:cand_nir_ext}.

\subsubsection{Far-infrared}

For the associations with sources of the far-infrared (FIR) domain, we used the Galactic Legacy Infrared Midplane Survey Extraordinaire (GLIMPSE) data\footnote{\url{https://cdsarc.cds.unistra.fr/viz-bin/cat/II/293}} collected by the Infrared Array Camera (IRAC) aboard the Spitzer Space Telescope \citep{2003PASP..115..953B}. The survey covers the entire region $|l| < 65 \degree$ and $|b| < 1\degree$ and goes up to $|b| < 4.2\degree$ in some regions, including our ROI. It detected sources between 3.6 and 8 $\mu$m providing 3$\sigma$ point-source sensitivity limit of 13--15.5 mag depending on the band. Each source of the catalog has a 0.3 arcsec error on its right ascension and declination, and we used $0.3$ arcsec as an estimation of the 1$\sigma$ uncertainty on the position of the GLIMPSE sources. Therefore, the value of $\theta_\mathrm{MW}$ used in Equation \ref{eq:pos_cm} is
\begin{equation}
        \theta_\mathrm{MW} = 2 \times 0.3\ \mathrm{arcsec.}
        \label{eq:crossm_glimpse}
\end{equation}

We note that the IRAC data were band merged with the 2MASS All-Sky Point Source Catalog and the potential 2MASS counterparts of GLIMPSE sources are recorded in the catalog, together with their K$_\mathrm{s}$ magnitude. The results of the X-FIR associations are presented in Section \ref{sec:cand_fir} as well as the resulting exclusions.

\subsubsection{NVSS}
\label{sec:cat_nvss}

The NRAO VLA Sky Survey (NVSS) has covered more than 80\% of the celestial sphere, including the Galactic center and produced a catalog of $1.7\times10^6$ sources stronger than 2.5 mJy at 1.4 GHz \citep{1998AJ....115.1693C}. We used
\begin{equation}
    \theta_\mathrm{MW} = \max(e_\mathrm{RA}, e_\mathrm{DEC})
\end{equation}
in Equation \ref{eq:pos_cm}, with $e_\mathrm{RA}$ ($e_\mathrm{DEC}$) the error on the right ascension (declination) of the NVSS source. Conclusions on the \textit{Chandra}-NVSS associations are presented in Section \ref{sec:cand_radio}. We note that a new VLA Sky Survey has recently started \citep{2020PASP..132c5001L}, but no catalog has been published yet.

\subsection{VLA L-band Galactic center mosaic}

Given the low point-source sensitivity of published radio data (section \ref{sec:cat_nvss}), we decided to analyze 28 hours of unpublished radio observations with the Karl G. Jansky Very Large Array (VLA) from project S9145 (PI: M. Kerr). We note that these data were used by \cite{2019ApJ...876...20H} to detect the very steep-spectrum polarized source C1748--2728, an MSP candidate in the GC. We checked that this source has no possible counterpart in the CSC. The VLA data were collected with 28 telescope pointings in a region around the Galactic center covering about 2$\degree$ in longitude and $3\degree$ in latitude (see Figure \ref{fig:cand_lb}). The total observation time was divided into twelve sessions of 2--4 hours and observed from January 9 to 19, 2017 in the VLA's A-configuration. Each session observed all 28 pointings for a duration of $\sim$200\,s per pointing in the L band using a pseudo-continuum configuration of 16 spectral windows, each having 64 x 1 MHz channels, to achieve a 1024 MHz instantaneous bandwidth and continuous frequency coverage over the 1.008 - 2.032 GHz range. The sources 3C 286 and J1751-2524 were included in each session for flux and gain calibration, respectively.

\subsubsection{Data reduction}
\label{sec:radio_data_red}

The VLA data were reduced using \texttt{CASA} \citep{2022PASP..134k4501C}, first by using the NRAO archive to apply the results of the VLA \texttt{CASA} calibration pipeline. The calibrated visibilities were split out by field center for each observing session and re-assembled into single-field data sets for further processing. A total of 62\% of these data were flagged due to observing overheads, radio interference, and poor calibration. Each of the 28 fields were imaged separately using \texttt{wsclean} \citep{2014MNRAS.444..606O}, convolved to a common resolution of 1.9 arcsec, and assembled into a linear mosaic.

\subsubsection{Data analysis}
\label{sec:radio_ana}

We looked for radio counterparts of the bulge MSP candidates by running a radio source detection algorithm, \texttt{PyBDSF}\footnote{\url{https://pybdsf.readthedocs.io/en/latest/}}, on the VLA mosaic. We selected squares of 100-pixel ($\sim$20\,arcsec) sides around the position of the $\sim$900 X-ray candidates covered by the VLA observations and ran \texttt{PyBDSF} on each of these regions. The algorithm calculates the background root mean square (RMS) in the reduced image, which goes up to $\sim$0.5\,mJy in the Galactic plane and on the borders of the mosaic and down to $\sim$0.02\,mJy outside of the plane. Then, it identifies pixels with flux larger than $t_\mathrm{pix}$ times the RMS and looks for islands of pixels with flux larger than $t_\mathrm{isl}$ times the RMS around them. $t_\mathrm{pix}$ and $t_\mathrm{isl}$ are threshold values defined either by the user (hard thresholds) or automatically computed with the false-detection-rate (FDR) method of \cite{2002AJ....123.1086H} implemented in \texttt{PyBDSF}. A Gaussian is then fit to each island. Certain sources (\ie, extended ones) sometimes require several Gaussians. We tested several configurations of the \texttt{PyBDSF} algorithm: i) a hard-threshold configuration with $t_\mathrm{isl} = 3$ and $t_\mathrm{pix} = 5$, ii) a second one with $t_\mathrm{isl} = 2$ and $t_\mathrm{pix} = 3,$ and iii) the FDR method. Configuration ii is motivated by the faintness of the sources that we are looking for, $\sim$100\,$\mu$Jy (see also Section \ref{sec:cand_radio} for a discussion on the radio emission of known MSPs).

This procedure identifies all radio sources in a 100-pixel-side square around \textit{Chandra} sources. As a second step, we applied the cross-match method described in Section \ref{sec:crossmatch} in order to find the radio counterparts of the MSP candidates. We neglected the error on the position of the radio sources, using Equation \ref{eq:sep_nir} again. A potential counterpart here would not exclude the pulsar nature. Therefore, unlike optical, UV, and FIR cross-matches, candidates with radio counterparts are tagged and remain in our selection, at least in a first stage, while candidates without radio counterparts are not dismissed (accept for association; see Section \ref{sec:crossmatch}). More details will be provided in Section \ref{sec:cand_radio}.

\section{Multiwavelength identification of \textit{Chandra} MSP candidates}
\label{sec:mw_msp}

\subsection{Optical constraints with \textit{Gaia}}
\label{sec:cand_opt}

\cite{2021PhRvD.104d3007B}, based on the work of \cite{antoniadis_gaia_2021}, explain that the optical counterparts of bulge MSPs should not be visible for \textit{Gaia}. This results from the combination of dim luminosities, strong absorption and large distances. Following the procedure described in Section \ref{sec:cat_opt}, we identified all MSP candidates with optical counterpart(s). From the 3158 sources selected in \cite{2021PhRvD.104d3007B}, 800 were found to have at least one \textit{Gaia} counterpart and were excluded. 2358 sources survived this selection. Positions are shown in Figure~\ref{fig:cand_lb}.

\new{In order to estimate the proportion of spurious associations, we shifted the positions in the \textit{Gaia} catalog by 5, 10, and 30 arcsec. The cross-match with no shift produces the maximum number of positive associations, with an excess from $\sim$8 to $\sim$16\% of positive matches with respect to the cross-match with shifts.}

\subsection{Ultraviolet constraints with XMM OM}
\label{sec:cand_uv}

Just as optical light, UV light from bulge MSPs is expected to be dim. Hence, we also disregarded any candidate with UV counterparts. Among the 3158 candidates from \cite{2021PhRvD.104d3007B}, we found 90 sources matching positively with XMM-OM sources. 30 of the MSP candidates matching with an XMM-OM source previously matched with an optical source (at least). Therefore, only 60 new objects were rejected from the candidate sample, which reduced to 2298 sources. Positions are again shown in Figure~\ref{fig:cand_lb}.

\subsection{Infrared constraints}
\label{sec:cand_ir}

Given the increasing sensitivity of IR instruments, the IR counterparts of MSPs in binary systems (\ie, the emission from their companion), although very faint, could be detectable. In order to distinguish them from other types of objects, we used the compact-object criterion.

\subsubsection{Compact object criterion}

 \citet{2012ApJ...756...27L} showed that compact objects have a high X-to-IR flux ratio, with
\begin{equation}
        \log_{10}(F_\mathrm{XMM}/F_\mathrm{K})\gtrsim0.5,
        \label{eq:co_crit}
\end{equation}
where $F_\mathrm{XMM}$ is the 0.2--12 keV absorbed X-ray flux seen by XMM,
\begin{equation}
        F_{\mathrm{K}} = 10^{-k_\mathrm{s}/2.5 - 6.95}
\end{equation}
is the infrared flux in the 2MASS K$_\mathrm{s}$ band in erg/cm$^2$/s, and $k_\mathrm{s}$ is the magnitude in this same band. In the following, we assume that the magnitude in the K$_\mathrm{s}$ band does not vary from one telescope to another, so Equation \ref{eq:co_crit} also applies to IR catalogs other than 2MASS. As for the X-ray flux, the \textit{Chandra} broad band being narrower than the XMM one mentioned above, we calculated, using the simulated MSP population of \cite{2021PhRvD.104d3007B}, that the XMM flux of (\textit{Chandra}-)detectable bulge MSPs should be at least equal to, and at most three times larger than, than the \textit{Chandra} broad-band flux. In the following, we assume
\begin{equation}
        F_\mathrm{XMM} = 3 \times \texttt{flux\_aper\_90\_b}
\end{equation}
for all MSP candidates, so we did not miss any compact object.\new{The compact-object criterion (Equation \ref{eq:co_crit}) was established from sources in the XMM catalog, which have an absorbing column density that is lower, on average, than the ones of bulge sources. However, the flux in band K$_\mathrm{s}$ is more affected by absorption than the X-ray flux of hard sources as our candidates. Therefore, the criterion also remains valid in the context of this work.}

A positive cross-match verifying the compact-object criterion would not exclude a pulsar nature, quite the contrary. Thus, if a candidate has a possible NIR counterpart for which the compact-object criterion is respected, the candidate is kept, regardless of other NIR cross-matches (accepting association). On the contrary, if counterparts are found and none of them respect the criterion, we exclude the candidate from the selection.

\subsubsection{Near-infrared point sources}
\label{sec:cand_nir}

A total of 1608 unique objects of the selection of \cite{2021PhRvD.104d3007B} matched positively with 2MASS, VVV, or UKIDSS sources, 1399 of which are too bright in the NIR to be MSPs. The other 209 sources respect the compact-object criterion, but 130 of them also match with UV or optical sources. We kept the 79 X-NIR sources with neither optical nor UV counterparts as compact-object (and therefore MSP) candidates for now. Among the $\sim$1400 ($\sim$1200) \textit{Chandra}-VVV (-UKIDSS) associations, $\sim$90 ($\sim$190) respect the compact-object criterion, including about 1/3 of matches with a unique potential association and 2/3 with at least two potential associations, including at least one respecting the criterion. There is of course some overlap between the associations made VVV and UKIDSS. The contribution of 2MASS sources to the compact-object candidates is negligible. In summary, 1608 candidates passed the NIR selection, and their positions are shown in Figure \ref{fig:cand_lb}.

\subsubsection{Extended objects}
\label{sec:cand_nir_ext}

We found a total of 72 MSP candidates from our conservative selection overlapping with five of the 2MASX sources in our ROI. Extended IR sources such as galaxies should be excluded from our MSP candidate selection, while extended IR sources like supernova remnants or pulsar wind nebulae should be kept. None of the 2MASX sources in our ROI are flagged as 'found in galaxy catalog'. Therefore, the 72 positive cross-matches cannot be disregarded.

\subsubsection{Background objects}
\label{sec:cand_fir}

We used the GLIMPSE data to identify reddened sources that we interpreted as background objects, that is, objects outside of the Milky Way and therefore outside of the Galactic bulge, or as obscured objects, which is not expected for bulge MSPs\footnote{Obscured objects typically have high-mass companions. While evolutionary tracks predict MSPs with high-mass companions to form systems such as the double pulsar PSR~J0737--3039, this phase should be short lived \citep[see, \eg,][which is also consistent with the fact that no MSP with a high-mass companion is known]{2004MNRAS.349..169D}. In addition, such high-mass stars are not typical in the bulge, making it even more unlikely to find such systems among our targets.}. We found 613 positive cross-matches between the \textit{Chandra} candidates of \cite{2021PhRvD.104d3007B} and GLIMPSE sources, which were all used to reduce the selection of MSP candidates. As previously mentioned, some GLIMPSE sources have been associated with 2MASS sources, and their K$_\mathrm{s}$ magnitude is provided in the catalog. The compact-object criterion was not verified for the 526 positive cross-matches with 2MASS associations. Out of the 79 compact-object candidates (VVV or UKIDSS counterpart) with neither UV nor optical counterpart, 22 matched positively with GLIMPSE sources and were excluded. 1422 candidates survived the FIR selection and their positions are shown in Figure \ref{fig:cand_lb}.

\begin{figure*}[ht!]
    \centering
    \includegraphics[width = 0.33\textwidth]{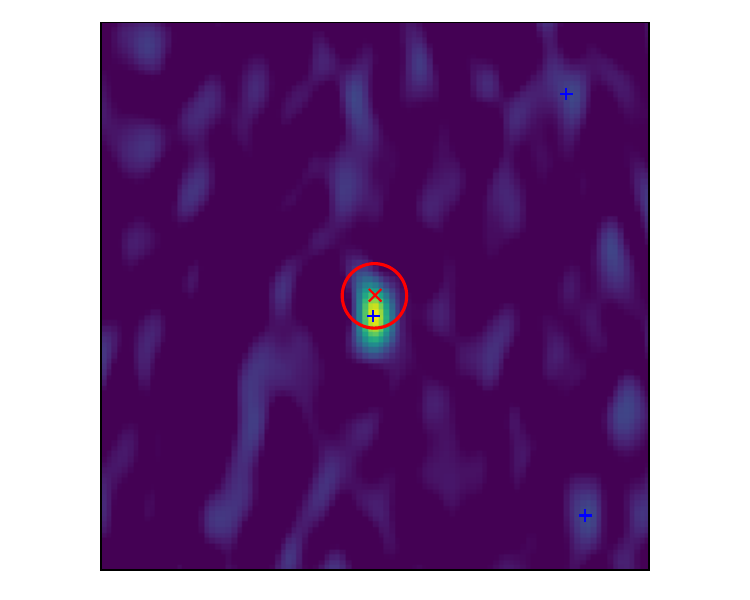}
    \includegraphics[width = 0.33\textwidth]{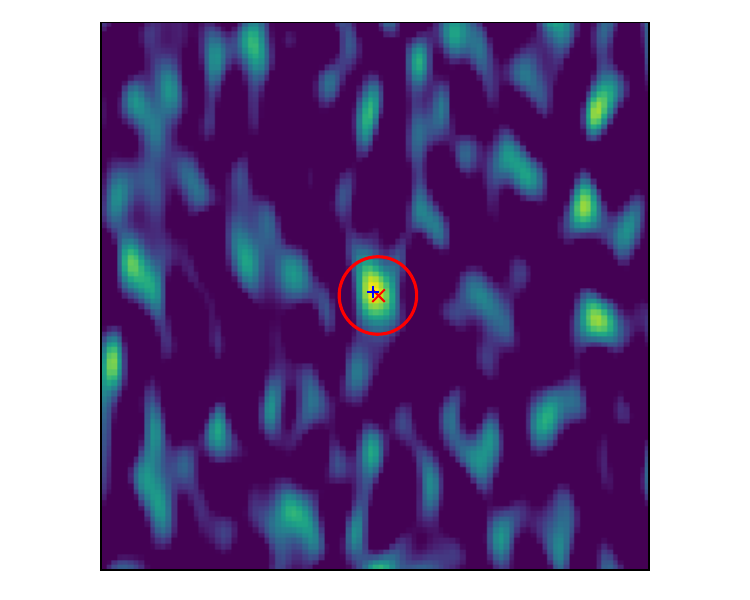}
    \includegraphics[width = 0.33\textwidth]{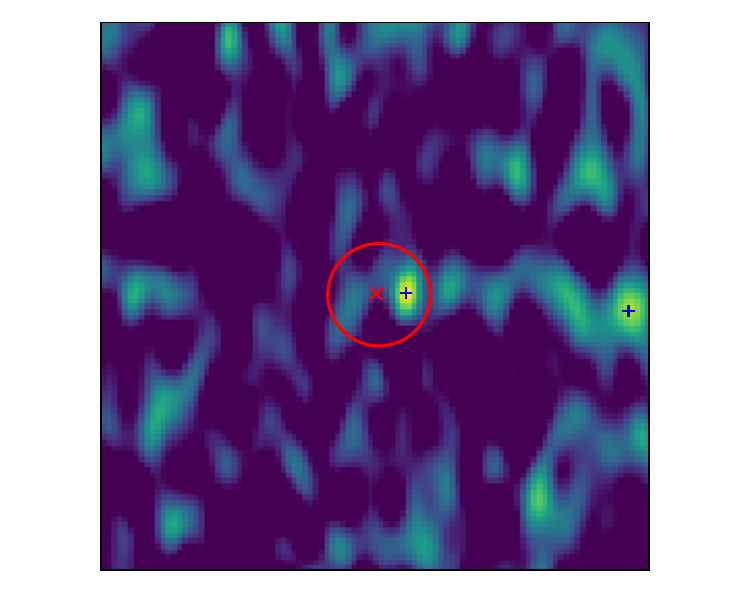}

    \includegraphics[width = 0.33\textwidth]{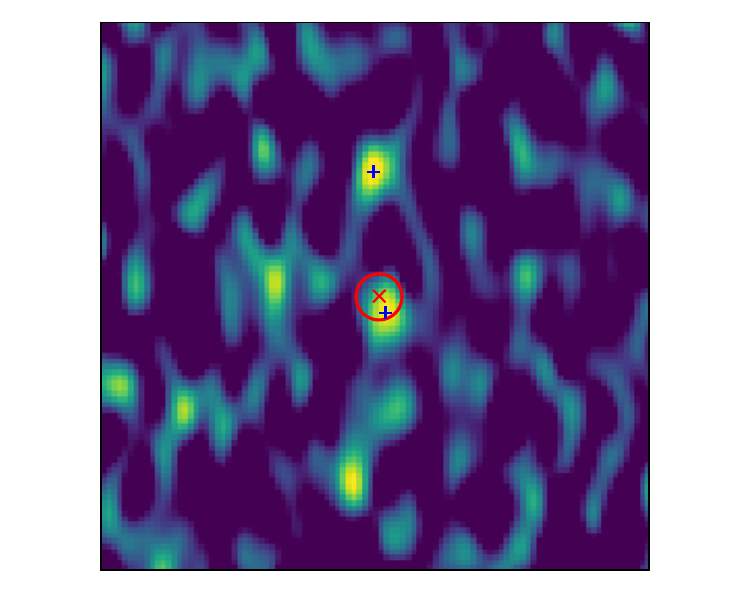}
    \includegraphics[width = 0.33\textwidth]{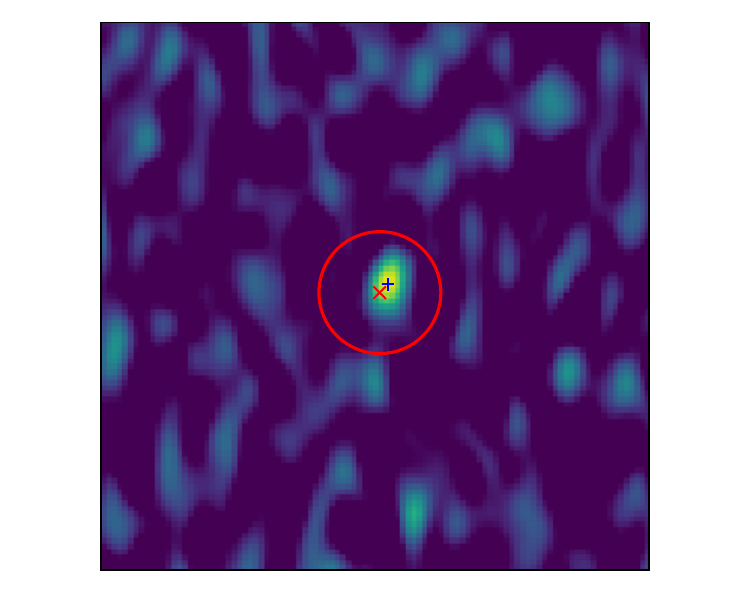}
        \caption{Association of \textit{Chandra} MSP candidates with VLA sources for our top priority candidates. The red crosses show the position of MSP candidates detected by \textit{Chandra}. The red circle around them has a radius of \texttt{err\_ellipse\_r0}. The blue crosses indicate the position of radio sources detected by \texttt{PyBDSF} (configuration ii). When the blue cross falls inside the red circle, we consider the X-ray and radio sources to be associated. The colored background shows the VLA mosaic data in units of $\mu$Jy/beam in $\sim$20\,arcsec-side squares around the X-ray sources. The color-scale is linear between 0 (purple) and the peak flux of the associated radio source quoted in Table \ref{tab:radio_exclusion} (yellow). From top left to bottom right, the \textit{Chandra} sources are 2CXO J173946.6--282913, 2CXO J174007.6--280708, 2CXO J174011.5--283221, 2CXO J174017.3--282843, and 2CXO J174053.7--275708. Other sources are shown in Figure \ref{fig:vla_ctp_appendix}.}
        \label{fig:vla_ctp}
\end{figure*}

\subsection{Radio associations and MSP-candidate identification}
\label{sec:cand_radio}

We found a total of 15 positive cross-matches between the \emph{Chandra}-MSP candidates and NVSS sources, with 1.4 GHz fluxes between 3.2 and 350.5 mJy. In the Australia Telescope National Facility (ATNF) pulsar catalog \citep{antoniadis_gaia_2021}, the largest 1.4 GHz pseudo-luminosity of an MSP is $\sim$170\,mJy\,kpc$^2$. Therefore, at 5.2 or 8.5\,kpc, this source would have a flux of 2.35 or 6.29 mJy. Alternatively, a flux of 3.2 mJy at 5.2 kpc corresponds to a pseudo-luminosity of $\sim$87\,mJy\,kpc$^2$, which is only reached by three MSPs in the ATNF pulsar catalog. The bulk of the MSPs in the ATNF pulsar catalog has a luminosity $\leq$10 mJy kpc$^2$,  and so a flux below 140 $\mu$Jy should they be located at the Galactic center. The radio fluxes of the \emph{Chandra}-NVSS sources are probably too high for MSPs in the bulge, but these candidates were not excluded at this point.

We found a total of 13 positive cross-matches between \emph{Chandra}-MSP candidates and VLA sources found with \texttt{PyBDSF}: eight are found in all \texttt{PyBDSF}-parameters configurations (i, ii, and FDR) and five in the loosest configuration only (ii). For consistency, all values (\eg, positions and fluxes) quoted in what follows refer to the results obtained in the loose configuration. To the best of our knowledge, at this point of our analysis, these sources only emit in radio and X-rays, except one that also has a UKIDSS counterpart that respects the compact-object criterion. We inspected all X-radio associations visually, thanks to plots similar to the ones shown in Figures \ref{fig:vla_ctp} and \ref{fig:vla_ctp_appendix}, where we show the 100-pixel-side squares around the \textit{Chandra} sources in the VLA mosaic, with the position of the X-ray and the radio sources. We conclude that two sources are actually noise, given the general aspect of the $\sim$20\,arcsec side radio image, that another one is an active galactic nucleus (confirming the statement of \cite{2012MNRAS.426..903P}) and that another one is resolved by the telescope (\ie, extended), while pulsars should be point-like, compact sources. We considered a source to be resolved when the width of the Gaussian fit to the island divided by the telescope synthesized beam width was larger than 1.5. These four sources are not further considered as radio MSP candidates.

In order to make sure that we did not miss any other potential counterpart, we search the positions of the nine remaining \emph{Chandra} sources in VizieR,\footnote{\url{https://vizier.cds.unistra.fr/viz-bin/VizieR}} which browses about 27~000 catalogs to find neighboring sources. The procedure reveals that two candidates have a potential optical counterpart in the Panoramic Survey Telescope and Rapid Response
System (Pan-STARRS) catalog\footnote{Had we used the Pan-STARRS catalog for the cross-matches, we would only have removed $\sim$70 additional sources.} \citep{2016arXiv161205560C}. Two other sources, the brightest of the nine ($>400$ $\mu$Jy), are found in quick-look images of the new VLA Sky Survey \citep{2020PASP..132c5001L}. 
We do not find any counterpart for four of the remaining sources, while the fifth is the one with an NIR (UKIDSS) counterpart that verifies the compact-object criterion.

In conclusion, we identify five \emph{Chandra}-MSP candidates with VLA association in the expected flux range ($\sim$100\,$\mu$Jy), seen for the first time in radio. These candidates are the most promising sources in our sample that will be worth following up with dedicated observing time. We emphasize that other sources in our selection remain interesting MSP candidates. Table \ref{tab:radio_exclusion} summarizes our selections; our five top-priority candidates are highlighted in bold and shown in Figure \ref{fig:vla_ctp}.

\begin{table*}[ht!]
        \caption{Summary of the radio cross-matches.}
        \centering
        \begin{tabular}{c|ccccc}
        \hline \hline
                2CXO name & Config. & Peak flux & RA & DEC & Comment \\ 
         & & $\mu$Jy/beam & deg & deg & \\ \hline 
        J173801.2--281352 & all & $82\pm13$ & --95.494942 & --28.231378 & Pan-STARRS counterpart\\
        \textbf{J173946.6--282913} & all & $181\pm13$ & --95.055515 & --28.487158 & \\
                J174000.6--274816 & all & $1683\pm108$ & --94.997272 & --27.80461 & AGN, NVSS counterpart\\
        J174000.7--274859 & all & $338\pm37$ & --94.99706 & --27.816485 & Pan-STARRS counterpart\\
        \textbf{J174007.6--280708} & ii & $45\pm13$ & --94.968179 & --28.119054 & \\
        \textbf{J174011.5--283221} & ii & $59\pm14$ & --94.952088 & --28.539374 & UKIDSS counterpart, compact-object candidate \\
        \textbf{J174017.3--282843} & ii & $43\pm14$ & --94.927984 & --28.478903 &\\
        \textbf{J174053.7--275708} & all & $69\pm12$ & --94.77617 & --27.952279 & \\
        J174309.3--292857 & all & $533\pm54$ & --94.21091 & --29.482736 & new VLA Sky Survey counterpart\\
        J174343.4--291358 & all & $406\pm33$ & --94.06908 & --29.232848 & new VLA Sky Survey counterpart\\
        J174602.4--284308 & ii & $108\pm22$ & --93.489942 & --28.718947 & noise\\
        J174616.3--284739 & ii & $202\pm56$ & --93.431595 & --28.794445 & noise\\
        J174810.0--285650 & all & $2341\pm152$ & --92.957553 & --28.947493 & not compact in size\\
                \hline
        \end{tabular}
        \tablefoot{From left to right, the columns give the name of the \textit{Chandra} source that matches with the radio source found by \texttt{PyBDSF}, the configuration of the algorithm (all: i and ii and FDR; see Section~\ref{sec:radio_ana}), the peak flux of the radio source and its $1\sigma$ error, its position in right ascension (RA) and declination (DEC), and comments on the radio detection and multiwavelength counterparts. Our five top-priority candidates, shown in Figure \ref{fig:vla_ctp}, are highlighted in bold (see main text for more details).}
        \label{tab:radio_exclusion}
\end{table*}

\section{Summary and discussion}
\label{sec:msp_mw_sumr}

In \cite{2021PhRvD.104d3007B}, we selected \textit{Chandra}-MSP candidates whose X-ray flux and spectral properties are consistent with those of bulge MSPs. Thanks to multiwavelength associations, we have now reduced this selection by a factor of $\sim\,$2 (from 3158 to 1422). To the best of our knowledge, sources that we kept in our selection emit in X-rays, but they are faint in the optical, UV, and IR, in agreement with the multiwavelength properties of known MSPs. Our selection includes more than 50 compact-object candidates with faint IR emission that could be MSPs in binary systems.

To reveal the most promising candidates we investigated VLA observations and revealed $\sim$10 candidates having a radio counterpart in the range of flux anticipated for MSPs in the Galactic bulge. Five of them (highlighted in bold in Table \ref{tab:radio_exclusion}) were selected for follow-up studies. Only four of them emit both in radio and X-rays, while the fifth one is also a compact-object candidate. Our global cross-match outcomes are summarized in Table \ref{tab:cand_summary}.

\subsection{Comparison with candidates in the inner parsecs}

Several groups recently published studies dedicated to populations of X-ray binaries \citep{2018Natur.556...70H, 2021ApJ...921..148M} and compact radio sources \citep{2020ApJ...905..173Z} close to the Galactic center, which could include pulsars and MSPs. We investigate whether there is any overlap between their selections and the one presented in this work.

\cite{2018Natur.556...70H} and \cite{2021ApJ...921..148M} performed a dedicated \textit{Chandra} data reduction to detect and investigate the nature of sources present in the central $\sim$$15~\rm pc$. None of the 12 sources they identified as persistent X-ray binaries are found in any of our selections. Seven of the X-ray binary candidates are not even found in the CSC, while the five remaining ones are flagged as extended and variable in the CSC. Therefore, they were excluded already from our MSP candidate list in \cite{2021PhRvD.104d3007B}. In this complex region, the CSC likely has several limitations, which can account for the discrepancies found.

\cite{2020ApJ...905..173Z} investigated the population of compact radio sources detected in a deep VLA image covering the central $\sim$40\,$\rm~pc$ and cross-matched their catalog with a dedicated \textit{Chandra} catalog of this complex region created by \citet{2018ApJS..235...26Z}. Two sources in our conservative selection could possibly be the X-ray counterparts of two compact radio sources identified by \cite{2020ApJ...905..173Z}, who did the reverse exercise of finding X-ray counterparts to their sources and were indeed successful with the two radio sources mentioned above, for which they also reported X-ray counterparts.
However, the two sources in our conservative selection were both disregarded as MSP candidates because one has a positive UV cross-match, while the second has a possible IR counterpart that does not respect the CO criterion. Other sources from \cite{2020ApJ...905..173Z} have possible counterparts in the CSC that are not in our selection because they are either extended, variable, or do not have the spectral shape expected for bulge MSPs. The remaining sources in their selection do not have possible X-ray counterparts in the CSC, and we emphasize that with our analysis, we cannot rule out their pulsar nature.

\begin{table}[t]
    \caption[Summary of the candidate selection]{Summary of the MSP candidate selection.}
        \centering
        \begin{tabular}{c|cc}
        \hline \hline
                Selection & Associations & Remaining candidates \\ \hline 
                Conservative & --- & \textbf{3158}\\
                Optical & 800 & 2358\\
                UV & 90 & 2298\\
                NIR & 1608 & 1483\\
        FIR & 613 & 1422\\ \hline
                Compact objects & 57 & ---\\
                VLA sources & 13 & ---\\
                \hline
        \end{tabular}
    \tablefoot{In the top section, we display the number of associations between the conservative MSP selection of \cite{2021PhRvD.104d3007B} and optical, UV, NIR, and FIR sources (middle column), and the number of remaining candidates after each selection (right column). In the bottom section, we highlight the candidates that survived all previous selections and that are either compact-object candidates or radio sources.}
        \label{tab:cand_summary}
\end{table}

\subsection{Nature of the remaining candidates}

The vast majority of the sources surviving the multiwavelength cross-matches performed in this work are only detected in X-rays.
We stress that, given the RMS of the VLA mosaic, the non-detection of the other MSP candidates in radio does not rule out their pulsar nature. In order to do so, and given the X-ray and radio properties of known MSPs, the RMS would need to improve by a factor of $\sim$2 at high latitudes, up to $\sim$100 in the Galactic plane, and even up to $\sim$500 (from $\sim$0.5\,mJy/beam to $\sim$1\,$\mu$Jy/beam) in the Galactic center region, where we have the highest density of candidates. We note that the RMS of the observations of \cite{2020ApJ...905..173Z} goes down to a few $\mu$Jy, which is still slightly too high to allow exclusions based on the absence of radio counterparts.

Observations of the inner part of our Galaxy were recently made with the MeerKAT telescope array, precursor of the Square Kilometer Array SKA \citep{2022ApJ...925..165H}, providing a total-intensity mosaic covering 6.5 square degrees, with an angular resolution of 4 arcsec (\ie, about twice the one in the VLA mosaic; Section \ref{sec:radio_data_red}). Moreover, the RMS in these MeerKAT data is either equivalent to or larger than the VLA mosaic one, depending on the position. The catalog of radio sources associated with the MeerKAT observations could nonetheless benefit our analysis once publicly available (Rammala et al. in prep.).

The overall number of candidates we could not exclude as MSP candidates ($\sim$1400) remains large compared with the predictions of X-ray detectable bulge MSPs from \cite{2021PhRvD.104d3007B} ($\sim$100). A fraction of magnetic cataclysmic variables, which are one of the dominant populations in our ROI \citep{2009ApJ...706..223H, 2011ApJS..194...18J}, could match the multiwavelength criteria we used to select our candidates; this probably represents the dominant contamination of our MSP candidate sample. An analysis of the X-ray properties of the population of candidates is needed to quantify this contamination and understand the population of sources with only X-ray emission that we discovered. This analysis will be presented in a separate publication.

\subsection{Conclusion}

With this work, we show that multiwavelength association is a successful way to reject and identify promising MSP candidates among unidentified X-ray sources, opening up new avenues for the search for MSPs. The absence of a deep radio imaging survey in our ROI, reaching the flux level expected for the bulk of bulge MSPs, was the limited factor in the selection of candidates for follow-up studies. The results of our timing observations are beyond the scope of the present work and will be presented in a dedicated publication.

\begin{acknowledgements}
      JB, FC and MC acknowledge financial support from the Programme National des Hautes Energies of CNRS/INSU with INP and IN2P3, co-funded by CEA and CNES, from the ‘Agence Nationale de la Recherche’, grant number ANR-19-CE310005-01 (PI: F. Calore), and from the Centre National d’Etudes Spatiales (CNES). This work supported by NASA under award number 80GSFC21M0002. Work at NRL is supported by NASA.
\end{acknowledgements}

%
%

\bibliographystyle{aa}
\bibliography{newbib}

\begin{appendix}

    \section{Other \textit{Chandra}-VLA cross-matches}

    In Section \ref{sec:cand_radio}, we explain how we selected our five top-priority candidates for radio follow-ups among the 13 positive \textit{Chandra}-VLA cross-matches. In this appendix, we show the VLA images of the eight candidates that were not selected, because of their size, nature, or spectral index for example.

    \begin{figure}[h]
        \centering
        
        \includegraphics[width = 0.24\textwidth]{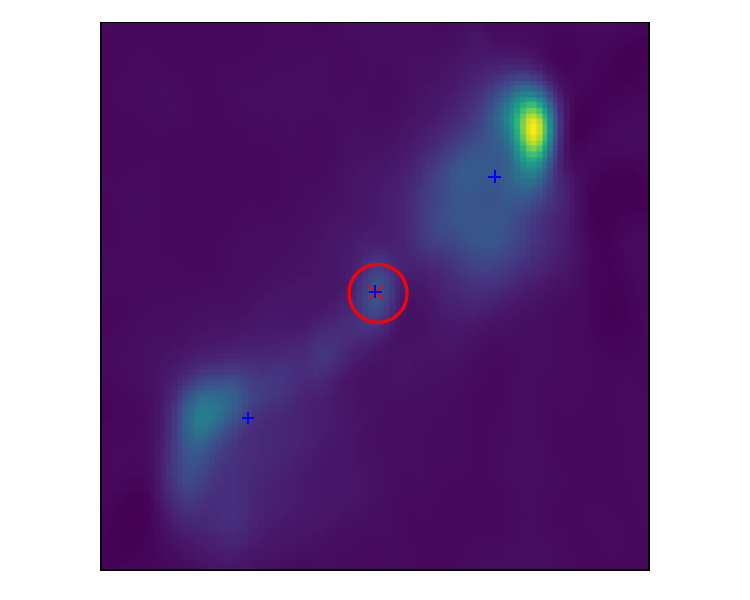}
        \includegraphics[width = 0.24\textwidth]{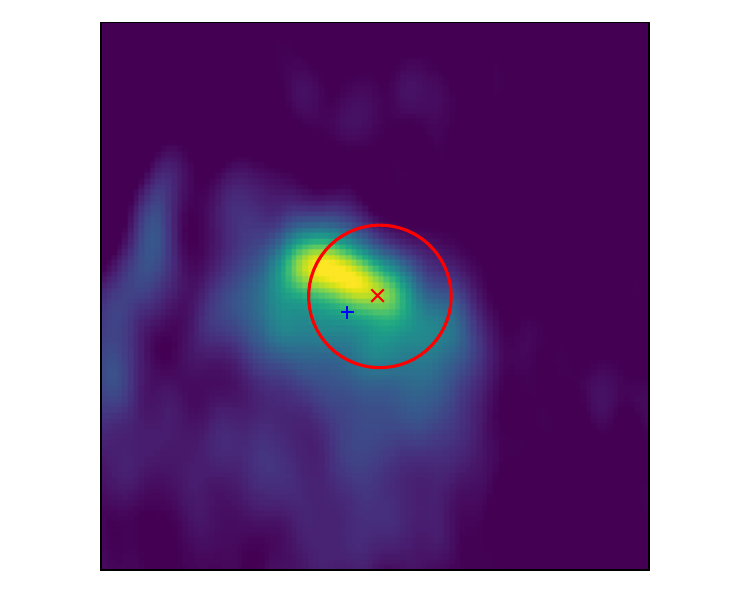}
        
        \includegraphics[width = 0.24\textwidth]{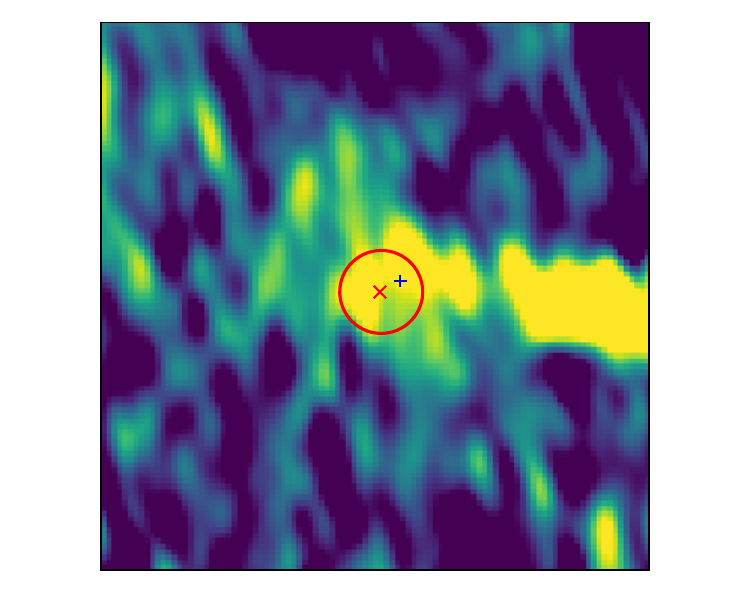}
        \includegraphics[width = 0.24\textwidth]{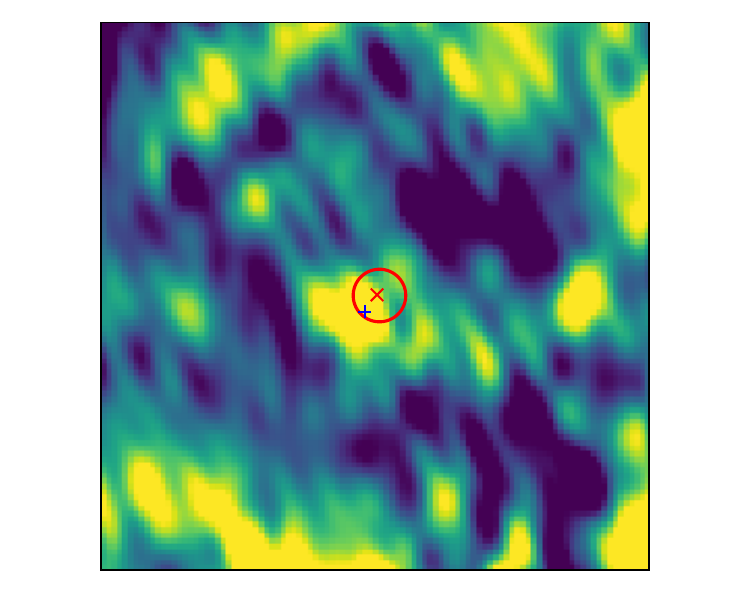}
        
        \includegraphics[width = 0.24\textwidth]{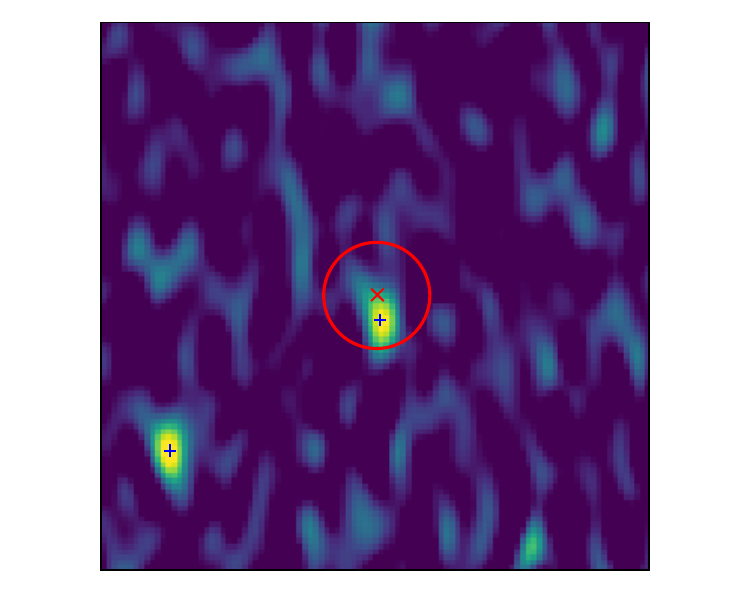}
        \includegraphics[width = 0.24\textwidth]{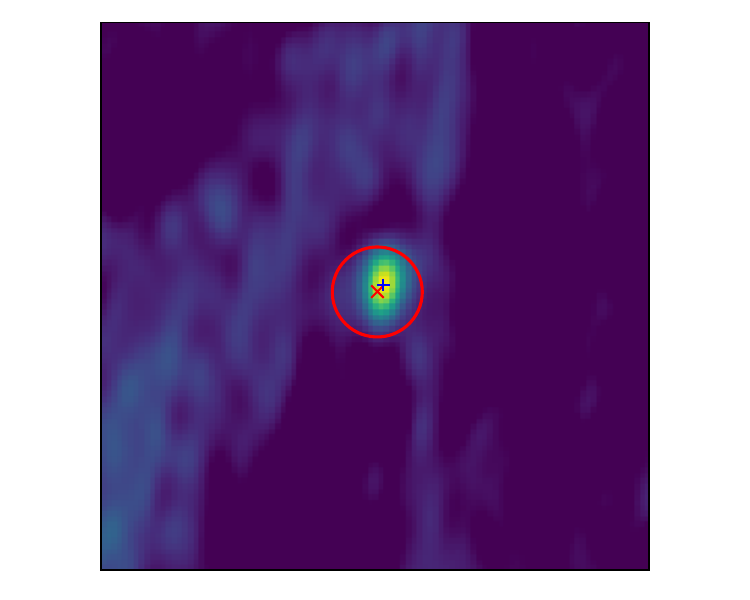}
        
        \includegraphics[width = 0.24\textwidth]{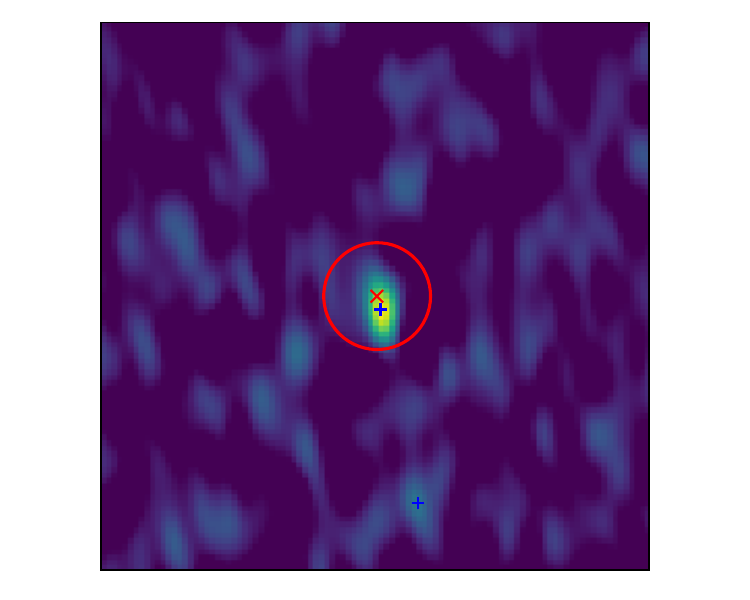}
        \includegraphics[width = 0.24\textwidth]{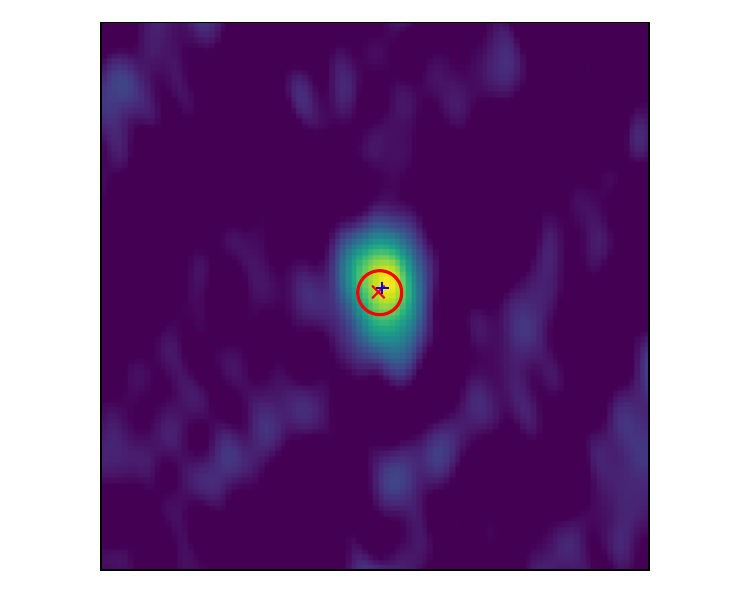}
        
        \caption{Same as Figure \ref{fig:vla_ctp}, but for (from top left to bottom right): 2CXO J174000.6--274816, 2CXO J174810.0--285650, 2CXO J174602.4--284308, 2CXO J174616.3--284739, 2CXO J173801.2--281352, 2CXO J174000.7--274859, 2CXO J174309.3--292857, and 2CXO J174343.4--291358. These sources were not selected for follow-up studies because they are either an AGN (top left), not compact (top right), noise (second line), have an optical counterpart (third line) or have a relatively strong and previously known radio emission (bottom line). The color scale for the AGN is linear between 0 and 8522 $\mu$Jy/beam, the maximal peak flux in the box.
        }
        \label{fig:vla_ctp_appendix}
    \end{figure}
    
\end{appendix}

\end{document}